\documentclass[twocolumn,superscriptaddress,preprintnumbers,amsmath,amssymb,pra]{revtex4-2}

\usepackage{graphicx}
\usepackage{dcolumn}
\usepackage{bm}
\usepackage{ifpdf}
\usepackage{epstopdf}
\usepackage{slashed}
\usepackage{amsfonts}
\usepackage{mathrsfs}
\usepackage{amssymb}
\usepackage{multirow}
\usepackage{CJK}
\usepackage{xcolor}
\usepackage{textcomp}
\usepackage{csquotes}

\def\lesssim{\ \raise.3ex\hbox{$<$}\kern-0.8em\lower.7ex\hbox{$\sim$}\ }
\def\gesim{\ \raise.3ex\hbox{$>$}\kern-0.8em\lower.7ex\hbox{$\sim$}\ }

\newcommand{\red}[1]{{{#1}}}

\begin{document}
\begin{CJK}{UTF8}{ipxm}
\preprint{RIKEN-iTHEMS-Report-22}

\title{Competition between pairing and tripling in one-dimensional fermions \\
with coexistent $s$- and $p$-wave interactions}

\author{Yixin Guo (郭一昕)}
\email{guoyixin1997@g.ecc.u-tokyo.ac.jp}
\affiliation{Department of Physics, Graduate School of Science, The University of Tokyo, Tokyo 113-0033, Japan}
\affiliation{RIKEN iTHEMS, Wako 351-0198, Japan}

\author{Hiroyuki Tajima (田島裕之)}
\email{htajima@g.ecc.u-tokyo.ac.jp}
\affiliation{Department of Physics, Graduate School of Science, The University of Tokyo, Tokyo 113-0033, Japan}

\date{\today}

\begin{abstract}
We theoretically investigate in-medium two- and three-body correlations in one-dimensional two-component Fermi gases with coexistent even-parity $s$-wave and odd-parity $p$-wave interactions.
We find the solutions of the stable in-medium three-body cluster states such as the Cooper triple by solving the corresponding in-medium variational equations.
We further feature a phase diagram consisting of the $s$- and $p$-wave Cooper pairing phases, and Cooper tripling phase, in a plane of $s$- and $p$-wave pairing strengths.
The Cooper tripling phase dominates over the pairing phases when both $s$- and $p$-wave interactions are moderately strong.
\end{abstract}

\maketitle

\section{Introduction}\label{sec:I}
In recent years, studies on superfluidity and superconductivity have attracted lots of focus in various fields.
Understanding the non-trivial states arising from the competition and the coexistence of more than two orders is one of the most important and challenging topics.
Along this direction,
the competing orders and clustering play an essential role for the realization of strongly correlated condensates~\cite{Mackenzie2003Rev.Mod.Phys.75.657--712,Stewart2011Rev.Mod.Phys.83.1589--1652,Stewart2017Adv.Phys.66.75--196,Xu2020J.Phys.:Condens.Matter32.343003}.
In condensed-matter systems, a fascinating example is 
an exotic state such as anapole superconductivity with competing even- and odd-parity pairing channels~\cite{Kanasugi2022Comm.Phys.5.1.} 
where its relevance for UTe$_2$ has been discussed recently.
In neutron stars and magnetars, $^1S_0$ and $^3P_2$ neutron superfluids~\cite{Takatsuka1993PTP.112.27} and moreover their coexistence~\cite{Yasui2020PhysRevC.101.055806} have also gathered considerable attention.
The $s$- and $p$-wave components of nuclear forces are also important for the formation of neutron-rich halo nuclei~\cite{Hammer2017}
and partially for tetraneutrons~\cite{marques2021quest,duer2022observation}.

A clean and controllable quantum system is suitable to investigate these unconventional states in a systematic manner.
An ultracold Fermi gas is regarded as an excellent platform for the study of many-body quantum systems. 
The remarkable feature of this atomic system is the controllable interaction through the Feshbach resonance~\cite{Chin2010Rev.Mod.Phys.82.1225--1286}.
In a three-dimensional $s$-wave superfluid Fermi gas, the pairing superfluid undergoes a crossover from a Bardeen-Cooper-Schrieffer (BCS) regime with weak-coupling Cooper pairs to a Bose-Einstein condensation (BEC) regime of tightly bound molecules~\cite{doi:10.1146/annurev-conmatphys-031113-133829,Strinati2018Phys.Rep.738.1--76,Ohashi2020Prog.Part.Nucl.Phys.111.103739}. 
The $p$-wave interaction is also tunable near the $p$-wave Feshbach resonance and $p$-wave Fermi gases have also been studied extensively towards the realization of $p$-wave Fermi superfluids~\cite{Ticknor2004Phys.Rev.A69.042712,Gurarie2005Phys.Rev.Lett.94.230403,Schunck2005Phys.Rev.A71.045601,Inada2008PhysRevLett.101.100401,Nakasuji2013PhysRevA.88.012710}.

\begin{figure}
  \includegraphics[width=9cm]{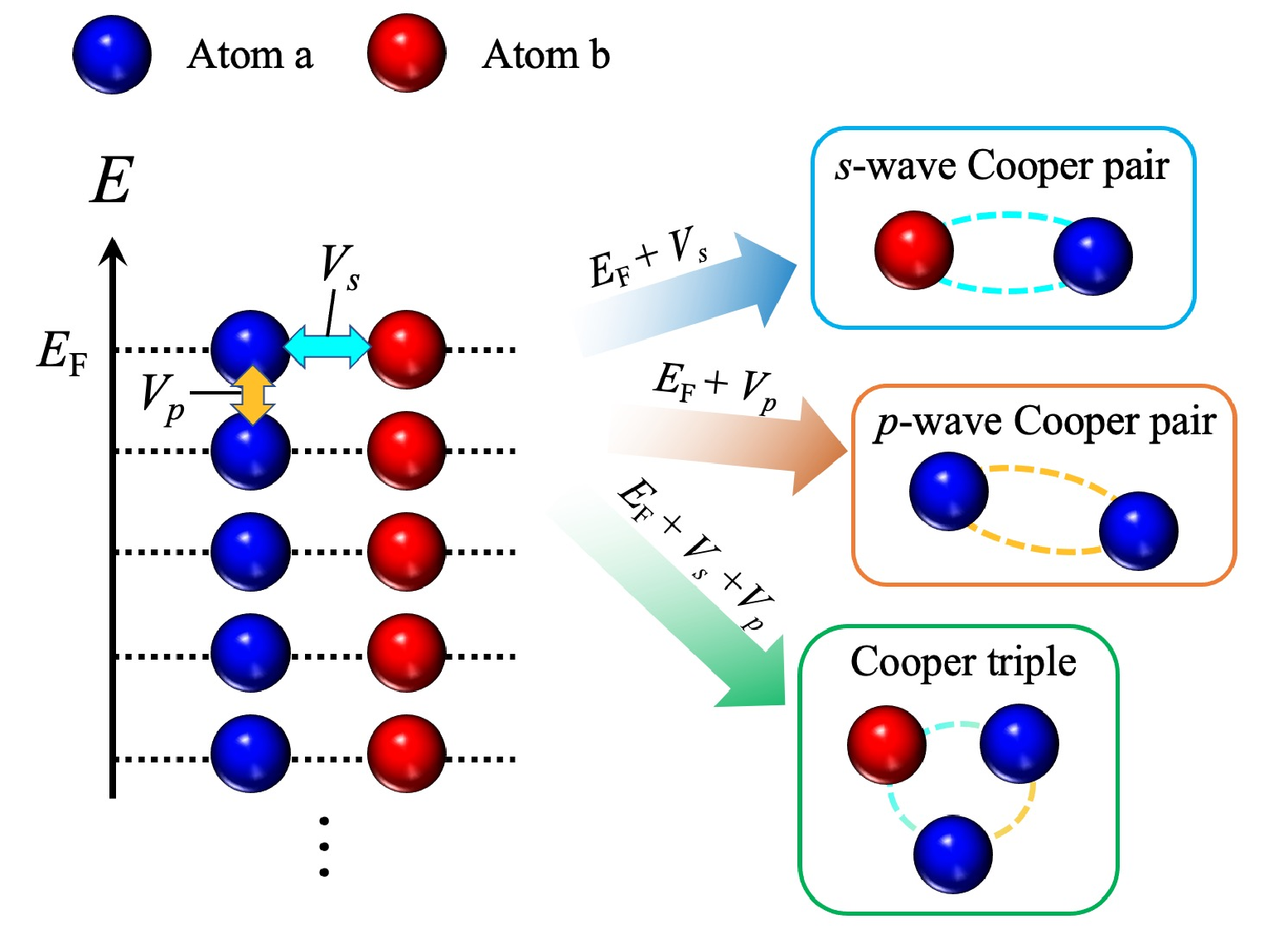}
  \caption{Schematic figure representing our model.
  We consider the degenerate two-component fermions (states $a$ and $b$ occupying the energy levels $E$ up to the Fermi energy $E_{\rm F}$) and the consequence of coexistent interspecies $s$-wave interaction $V_s$ and intraspecies $p$-wave interactions $V_p$ (acting on two identical fermions in the $a$ state),
  which lead to the Cooper instabilities towards the $s$- and $p$-wave Cooper pairs, and the Cooper triple.
  }\label{fig:0}
\end{figure}

As a step forward, it is also exciting to figure out the properties in a system with both $s$- and $p$-wave interactions as shown in Fig.~\ref{fig:0}.
In addition to the interdisciplinary potential interests in condensed-matter and nuclear physics,
such a situation can be realized in cold atomic systems.
Indeed, Fermi superfluids with hybridized $s$- and $p$-wave pairings, which can be realized in two-component $^{40}$K Fermi gases near the overlapped $s$- and $p$-wave magnetic Feshbach resonances~\cite{Chin2010Rev.Mod.Phys.82.1225--1286,Regal2003PhysRevLett.90.053201},
have been studied theoretically in 
Ref.~\cite{Zhou2017ScienceChina.60.12}.
A Borromean trimer is also predicted in a three-dimensional mass-imbalanced mixture with the hybridized interactions~\cite{10.21468/SciPostPhys.12.6.185}.
Moreover, an emergent $s$-wave interaction in quasi-one-dimensional Fermi gases near the $p$-wave resonance has been reported experimentally~\cite{Jackson2022emergent}.

While superconductors/superfluids with $s$- and $p$-wave Cooper pairs have been studied extensively,
such a clustering associated with the Cooper instability is not necessarily limited to Cooper pairs but may involve more-than-three-body bound states in the presence of both $s$- and $p$-wave interactions.
To investigate the three-body clustering in quantum many-body systems, we need to consider the in-medium three-body problem.
For such a purpose, the generalized Cooper problem has been further applied to cluster states such as Cooper triples~\cite{Niemann2012Phys.Rev.A86.013628,Kirk2017Phys.Rev.A96.053614,Akagami2021Phys.Rev.A104.L041302,Tajima2021Phys.Rev.A104.053328,Tajima2022Phys.Rev.Research4.L012021} and even Cooper quartets~\cite{Roepke1998Phys.Rev.Lett.80.3177--3180,Sandulescu2012Phys.Rev.C85.061303,Baran2020Phys.Lett.B805.135462,Guo2022Phys.Rev.C105.024317,Guo2022Phys.Rev.Research4.023152}.
\red{These Cooper clusters may exhibit nontrivial many-body properties distinct from conventional superconductors. 
While the many-body properties of the Cooper triple phase are still elusive, several theoretical proposals about this point have been reported in Refs.~\cite{Niemann2012Phys.Rev.A86.013628,Kirk2017Phys.Rev.A96.053614,Akagami2021Phys.Rev.A104.L041302,Tajima2021Phys.Rev.A104.053328,Tajima2022Phys.Rev.Research4.L012021}, where the quantum phase transition from the BCS superfluid phase to the Cooper triple phase have been discussed.
Moreover, the three-body loss would be an experimental signature for the emergence of Cooper triples as in the case of Efimov effects in cold atomic systems~\cite{Tajima2021Phys.Rev.A104.053328,Naidon2017Rep.Prog.Phys.80.056001}.}
The investigation of the fate of such higher-order clustering is also a fascinating topic in various fields.
These approaches are useful for a further understanding of the many-body ground states. 

Another advantage of cold atomic systems is
a controllable dimensionality associated with the trap potential~\cite{Bloch2008Rev.Mod.Phys.80.885--964}.
The realization of a low-dimensional system near the Feshbach resonance leads to the enhanced pairing effects known as confinement-induced resonance~\cite{Olshanii1998PhysRevLett.81.938,Bergeman2003PhysRevLett.91.163201,Moritz2005PhysRevLett.94.210401}.
Moreover, the stability against the three-body loss
in a one-dimensional fermionic system near the $p$-wave Feshbach resonance has been predicted theoretically~\cite{Zhou2017PhysRevA.96.030701,Fonta2020PhysRevA.102.043319} and recently
the several experimental groups have performed loss measurements in this system~\cite{Chang2020PhysRevLett.125.263402,marcum2020suppression,Jackson2022emergent}.
Apart from these backgrounds associated with cold atomic experiments, these atomic systems are of interest as quantum simulators of different low-dimensional condensed-matter and nuclear systems.
Indeed, one-dimensional superconductors have attracted much attention in condensed-matter physics~\cite{altomare2013one}.
In nuclear physics, the low-dimensional systems are considered as benchmark models~\cite{Alexandrou1989PhysRevC.39.1076} or some specific configurations such as two-neutron halo nuclei in a one-dimensional mean field~\cite{Hagino2010}.

In this paper, we investigate a one-dimensional fermionic system with both $s$- and $p$-wave interactions schematically shown in Fig.~\ref{fig:0}.
For simplicity, we consider the spin- and mass-balanced case in this work.
By solving the in-medium three-body equation derived from the variational principle, we show the solutions of the stable in-medium three-body cluster state (such as a Cooper triple) in the present system.
Accordingly, we also show a phase diagram consisting of $s$- and $p$-wave pairing states, and the Cooper triple states.
Our result can be tested in $^{40}$K Fermi gases near the overlapped $s$- and $p$-wave resonances around $B=200$ G~\cite{Zhou2017ScienceChina.60.12}.
In addition, our model with both $s$- and $p$-wave interactions is similar to the recent experiment of $^{40}$K Fermi gases~\cite{Jackson2022emergent}, where an $s$-wave interaction emerges near the $p$-wave Feshbach resonance due to the quasi-one-dimensionality.

This paper is organized as follows.
The theoretical framework is presented in Sec.~\ref{sec:II}, where we show the Hamiltonian for a one-dimensional two-component Fermi gas with both $s$- and $p$-wave interactions. 
We apply a variational method for in-medium three-body states on top of the Fermi sea to this model. 
In Sec.~\ref{sec:III}, we show our numerical results for the in-medium bound states and the ground-state phase diagram.
Finally, a summary and perspectives will be given in Sec.~\ref{sec:IV}.
In the following, we take $\hbar=c=k_{\rm B}=1$.
The system size is taken to be a unit.

\section{Theoretical Framework}\label{sec:II}

We consider one-dimensional two-component fermions with coexistent $s$- and $p$-wave interactions.
The Hamiltonian of such a system reads
\begin{align}
    H=\,&K+V_s+V_p,\\
    K=\,&\sum_{k}\left(\xi_{k,a}c_{k,a}^\dag c_{k,a}+\xi_{k,b}c_{k,b}^\dag c_{k,b}\right),\\
    V_p=\,&\frac{U_{p}}{2}\sum_{p,p',q}pp'c_{p+q/2,a}^\dag c_{-p+q/2,a}^\dag c_{-p'+q/2,a} c_{p'+q/2,a},\\
    V_s=\,&U_s\sum_{p,p',q}c_{p+q/2,a}^\dag c_{-p+q/2,b}^\dag c_{-p'+q/2,b} c_{p'+q/2,a},
\end{align}
where $c_{k,a}^{(\dag)}$ and $c_{k,b}^{(\dag)}$ represent the annihilation (creation) operators of the two-component fermions with the states $a$ and $b$ (e. g., hyperfine states and spins), respectively; here $\xi_{k,i}=k^2/(2m_i)-\mu_i$ ($i=a,b$) in the kinetic term $K$ is the single-particle energy with a momentum $k$, an atomic mass $m_i$, and chemical potential $\mu_i$.
For simplicity, we consider the mass- and spin-balanced system as $m\equiv m_a=m_b$ and $\mu\equiv \mu_a=\mu_b$.
In the generalized Cooper problems, we take $\mu=E_{\textrm{F}}$ where $E_{\textrm{F}}$ is the Fermi energy.
$V_p$ represents the short-range $p$-wave two-body interaction with a coupling constant $U_p$, and $V_s$ corresponds to the $s$-wave two-body interaction with a coupling constant $U_s$.
Here, the contact couplings $U_s$ and $U_p$ can be renormalized by introducing the $s$-wave and $p$-wave scattering lengths~\cite{Guan2013RevModPhys.85.1633,Cui2016PhysRevA.94.043636} as
\begin{align}
    U_s=-\frac{2}{ma_s},
\end{align}
and
\begin{align}
    \frac{m}{2a_p}=\frac{1}{U_p}+\sum_{p}\frac{p^2}{2\varepsilon_p}
\end{align}
with $\varepsilon_{p}=p^2/(2m)$,
respectively.
Since we are interested in an attractive $s$-wave interaction,
the positive $s$-wave scattering length $a_s>0$ is taken.
The $p$-wave scattering length can be taken as both positive and negative values, and $1/(k_{\rm F}a_p)=0$ corresponds to the $p$-wave unitarity~\cite{Cui2016PhysRevA.94.043636,Tajima2021PhysRevA.104.023319}.

For convenience, here we further introduce the pair operators as
\begin{align}
    S_{p,q}^\dag = c_{p+q/2,a}^\dag c_{-p+q/2,b}^\dag,
\end{align}
and
\begin{align}
    P_{p,q}^\dag = c_{p+q/2,a}^\dag c_{-p+q/2,a}^\dag.
\end{align}
Correspondingly, the Hamiltonian can be rewritten as
\begin{align}
    H=\,&\sum_{k}\left(\xi_{k,a}c_{k,a}^\dag c_{k,a}+\xi_{k,b}c_{k,b}^\dag c_{k,b}\right)\nonumber \\
    &+\frac{U_{p}}{2}\sum_{p,p',q}pp'P_{p,q}^\dag P_{p',q}
    +U_s\sum_{p,p',q}S_{p,q}^\dag S_{p',q}.
\end{align}
The trial wave function for the in-medium three-body state is adopted as
\begin{align}
    |\Psi_{\rm CT}\rangle=\,&\sum_{p,q}
    \theta(|p+q/2|-k_{\rm F})\theta(|-p+q/2|-k_{\rm F})\nonumber\\
   & \times\theta(|-q|-k_{\rm F})
    \Omega_{p,q}F_{p,q}^\dag|{\rm FS}\rangle\cr
    &\equiv \sum_{p,q}'\Omega_{p,q}F_{p,q}^\dag|{\rm FS}\rangle,
\end{align}
where 
\begin{align}
    F_{p,q}^\dag=c_{p+q/2,a}^\dag c_{-p+q/2,a}^\dag c_{-q,b}^\dag
\end{align}
creates a triple above the Fermi sea, and $|{\rm FS}\rangle$ denotes the Fermi sea.
In addition, here $\displaystyle\sum_{p_1,p_2,\cdots}'$ is adopted to denote the the momentum summation restricted by the Fermi surface for convenience. 
The step functions associated with the Fermi surface will be recovered when we evaluate the momentum summation numerically.
By minimizing the ground-state energy based on the variational principle, the variational parameter $\Omega_{p,q}$ will be determined correspondingly, and it is easy to find $\Omega_{p,q}=-\Omega_{-p,q}$.
Based on the fact that $q$ describes the relative momentum between a $p$-wave pair of two identical fermions in state $a$ and a fermion in state $b$,  we assume even parity between them as $\Omega_{p,q}=\Omega_{p,-q}$.

The expectation values for the kinetic and interaction parts are obtained as
\begin{align}
&\langle\Psi_{\rm CT}|K|\Psi_{\rm CT}\rangle\nonumber\\
=\,&
\sum_{p,q,p',q',k}'\left(
    \xi_{k,a}\Omega_{p,q}^*
    \Omega_{p',q'}
    \langle{\rm FS}|F_{p,q}
    \red{
    c_{k,a}^\dag 
    c_{k,a}}
    F_{p',q'}^\dag|{\rm FS}\rangle\right.\nonumber\\
    &\left.+
    \xi_{k,b}\Omega_{p,q}^*
    \Omega_{p',q'}
    \langle{\rm FS}|F_{p,q}
    \red{c_{k,b}^\dag c_{k,b}}
    F_{p',q'}^\dag|{\rm FS}\rangle\right)\nonumber\\
=\,&2\sum_{p,q}'\left(
\xi_{p+q/2,a}+\xi_{-p+q/2,a}
+\xi_{-q,b}\right)|\Omega_{p,q}|^2,
\end{align}
and
\begin{align}
    \langle\Psi_{\rm CT}|V_s|\Psi_{\rm CT}\rangle
    =\,&U_s\sum_{k,k',Q,p,q,p',q'}'
    \Omega_{p,q}^*\Omega_{p',q'}\nonumber\\
    &\times\langle{\rm FS}|F_{p,q}S_{k,Q}^\dag S_{k',Q}F_{p',q'}^\dag|{\rm FS}\rangle\nonumber\\
    =\,&2U_s\sum_{p,q,q'}'\Omega_{p,q}^*
    \left(\Omega_{p+q/2-q'/2,q'}\right.\nonumber\\
    &\left.+\Omega_{p-q/2+q'/2,q'}\right),\\
    \langle\Psi_{\rm CT}|V_p|\Psi_{\rm CT}\rangle
    =\,&\frac{U_p}{2}\sum_{k,k',Q,p,q,p',q'}'
    kk'\Omega_{p,q}^*\Omega_{p',q'}\nonumber\\
    &\times\langle{\rm FS}|F_{p,q}P_{k,Q}^\dag P_{k',Q}F_{p',q'}^\dag|{\rm FS}\rangle\nonumber\\
    =\,&2U_p
    \sum_{p,q,p'}'pp'\Omega_{p,q}^*\Omega_{p',q},
\end{align}
respectively.

Furthermore, from the variational principle, we obtain 
\begin{align}
    \frac{\delta}{\delta \Omega_{{p},{q}}^*}\langle\Psi_{\rm CT}|(H-E_3)|\Psi_{\rm CT}\rangle=0,
\end{align}
where $E_3$ is the ground-state energy of a Cooper triple state.
Consequently, the variational equation reads
\begin{align}
    &(\xi_{p+q/2,a}+\xi_{-p+q/2,a}+\xi_{-q,b}-E_3)
    \Omega_{p,q}+U_p\sum_{p'}'pp'\Omega_{p',q}\nonumber\\
    &+U_s\sum_{q'}'(\Omega_{p+q/2-q'/2,q'}
    +\Omega_{p-q/2+q'/2,q'})=0.
\end{align}
In order to simplify the further derivations, here we introduce
\begin{align}
    \Gamma_p(q)=U_p\sum_{p'}'p'\Omega_{p',q},
\end{align}
and
\begin{align}
    \Gamma_s(k)=U_s\sum_{q'}'\Omega_{k+q'/2,q'}.
\end{align}
One can further find that
\begin{align}
    \Gamma_p(q)=\Gamma_{p}(-q),
    \quad
    \Gamma_s(k)=-\Gamma_s(-k).
\end{align}

Consequently, the variational equation can be recast into
\begin{align}
   & (\xi_{p+q/2,a}+\xi_{-p+q/2,a}+\xi_{-q,b}-E_3)
    \Omega_{p,q}+p\Gamma_p(q)\nonumber\\
    &+\Gamma_s(p-q/2)-\Gamma_s(-p-q/2)=0.
\end{align}
Correspondingly, one has the in-medium three-body equations for $\Gamma_p(q)$ and $\Gamma_s(k)$ as
\begin{align}\label{gammap}
    &\Gamma_p(q)
    \left[
    1+U_p\sum_{p}'
    \frac{
    p^2}{\xi_{p+q/2,a}+\xi_{-p+q/2,a}+\xi_{-q,b}-E}
    \right]\cr
    =\,&-U_p\sum_{p}'p\frac{\Gamma_s(p-q/2)-\Gamma_s(-p-q/2)
    }{\xi_{p+q/2,a}+\xi_{-p+q/2,a}+\xi_{-q,b}-E},
\end{align}
and 
\begin{align}\label{gammas}
    &\Gamma_{s}(k)
    \left[   1+U_s\sum_{q}'\frac{1}{\xi_{k+q,a}+\xi_{-k,a}+\xi_{-q,b}-E}
    \right]\cr
    =\,&-U_s\sum_{q}'\frac{(k+q/2)\Gamma_p(q)-\Gamma_s(-k-q)
    }{\xi_{k+q,a}+\xi_{-k,a}+\xi_{-q,b}-E},
\end{align}
respectively.

For comparison, we also calculate $s$- and $p$-wave Cooper pairing energies $E_{2,s}$ and $E_{2,p}$,
which can be obtained from the in-medium two-body equations for the $s$-wave pairing~\cite{Ohashi2020Prog.Part.Nucl.Phys.111.103739} 
\begin{align}
\label{eq:e2s}
1+U_s\sum_{q}'\frac{1}{\xi_{q,a}+\xi_{q,b}-E_{2,s}}=0,
\end{align}
and for the $p$-wave pairing~\cite{guo2022stability}
\begin{align}
\label{eq:e2p}
    1+U_p\sum_{p}'\frac{p^2}{\xi_{p,a}+\xi_{-p,a}-E_{2,p}}=0,
\end{align}
respectively.

\red{
At the end of this section,
we note that our variational approach can be used to describe the BCS-BEC crossover and its three-body counterpart qualitatively~\cite{Tajima2021PhysRevA.104.023319}.
It is well known that the three-dimensional BCS-BEC crossover has been studied by the BCS-Leggett mean-field theory~\cite{doi:10.1146/annurev-conmatphys-031113-133829,Strinati2018Phys.Rep.738.1--76,Ohashi2020Prog.Part.Nucl.Phys.111.103739}, where the deviation of the chemical potential from the Fermi energy is allowed. 
In the BCS-Leggett state (i.e., a BCS-like wavefunction), as the attraction increases, the chemical potential becomes different from the Fermi energy and turns into a negative one in the BEC regime. 
Consequently, it can properly describe the molecule formation in the BEC limit regardless of its mean-field framework. 
Similarly, the variational wave function of Cooper problems for the calculation of the clustering energy can also reproduce both the Cooper pairing in the weak-coupling BCS limit and the molecular formation in the strong-coupling BEC limit. 
Moreover, the generalized Cooper problem investigated here for the three-body sector can also describe both the Cooper triples in the weak-coupling regime and the trimer formation in the strong-coupling regime~\cite{Niemann2012Phys.Rev.A86.013628,Tajima2021PhysRevA.104.023319}. 
In this regard, the variational approach adopted here based on the extension of the few-body problem enables us to study the weak- and strong-coupling regimes in a unified manner.
}

\section{Results and discussion}\label{sec:III}

\begin{figure}
  \includegraphics[width=0.45\textwidth]{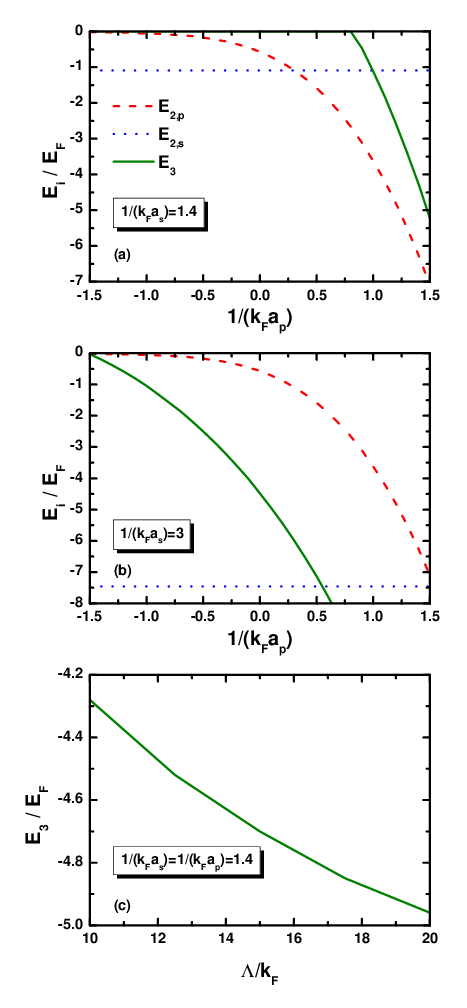}
  \caption{ 
  Calculated in-medium three-body energy $E_3$ and $s(p)$-wave pairing energy $E_{2,s(p)}$ as functions of the inverse $p$-wave scattering length $1/(k_{\rm F}a_p)$ at (a) $1/(k_{\rm F}a_s)=1.4$ and (b) $1/(k_{\rm F}a_s)=3$.
  The panel (c) shows the cutoff dependence of $E_3$ at $1/(k_{\rm F}a_p)=1/(k_{\rm F}a_s)=1.4$.
  }\label{fig:1}
\end{figure}

Figure~\ref{fig:1} shows the numerical results of the in-medium three-body energy $E_3$ obtained by solving Eqs.~\eqref{gammap} and \eqref{gammas},
where the momentum cutoff $\Lambda=10k_{\rm F}$ is used.
One can find that in general, at certain $1/(k_{\rm F}a_p)$, $|E_3|$ becomes larger with an increase of $1/(k_{\rm F}a_s)$.
Correspondingly, by adopting $1/(k_{\rm F}a_s)=3$, the solutions to the in-medium three-body equations~\eqref{gammap} and \eqref{gammas} can be even found in the weak-coupling side, which is shown in of Fig.~\ref{fig:1}(b).

In addition, $E_3$ also exhibits a cutoff dependence as shown of Fig.~\ref{fig:1}(c).
$|E_3|$ tends to increase with increasing $\Lambda$ as found in the one-dimensional system with a $p$-wave interaction and the three-body coupling~\cite{Sekino2021Phys.Rev.A103.043307,guo2022stability}.
\red{
The physical origin of such a UV cutoff is associated with the short-range properties of the interaction (e.g., effective range, short-range repulsion)~\cite{Chin2010Rev.Mod.Phys.82.1225--1286}.
On the other hand, a non-mean-field correlational collapse called the Thomas collapse~\cite{Thomas1935Phys.Rev.47.903--909}, where the short-range attractive interaction induces a deep three-body bound state with a large binding energy proportional to $\Lambda^2$, appears in some of the three-body problems.
It is in stark contrast to the two-body problem and the associated Cooper pairing. 
In the present case, even only with the $p$-wave interaction, the three-body integral equation shows an explicit ultraviolet divergence~\cite{Sekino2021Phys.Rev.A103.043307,guo2022stability}, regardless of the absence of three-body bound states. 
Since our approach incorporates such cutoff-dependent three-body properties with the Pauli-blocking effect,
our numerical results also exhibit a cutoff dependence.
We also note that
the three-body parameter~\cite{Naidon2017Rep.Prog.Phys.80.056001} can be introduced to regularize the zero range theory of three particles, e.g., the three-body parameter $\kappa$ as $E_3=\kappa^2/m$.
In the three-dimensional case~\cite{Tajima2021Phys.Rev.A104.053328}, the finite cutoff plays a role of such a three-body parameter, which is similar to the present work.
}

While the Cooper triple and the trimer states are qualitatively distinguished in Ref.~\cite{Tajima2021Phys.Rev.A104.053328},
in this paper we do not go into details about their differences because these two states have no distinct boundaries as in the case of the BCS-BEC crossover~\cite{doi:10.1146/annurev-conmatphys-031113-133829,Strinati2018Phys.Rep.738.1--76,Ohashi2020Prog.Part.Nucl.Phys.111.103739}, and moreover such boundaries quantitatively depend on $\Lambda$ in our model.

To see a competition between pairing and tripling, we also plot the energies of $s$- and $p$-wave Cooper pairs $E_{2,s}$ and $E_{2,p}$ obtained from Eqs.~\eqref{eq:e2s} and \eqref{eq:e2p} in Figs.~\ref{fig:1}(a) and \ref{fig:1}(b).
These two-body energies can be regarded as the Cooper pairing binding energy with a remaining unpaired fermion on the Fermi sea, and hence they can be directly compared with $E_3$ as the energy gains of each cluster.
Because $E_{2,s}$ does not depend on $1/(k_{\rm F}a_p)$ through Eq.~\eqref{eq:e2s} in the Cooper problem,
$E_{2,s}$ remains a constant at fixed $1/(k_{\rm F}a_s)$ in Figs.~\ref{fig:1}(a) or \ref{fig:1}(b).
$E_{2,p}$ is also independent of $1/(k_{\rm F}a_s)$ and therefore $E_{2,p}$ shown in Figs.~\ref{fig:1}(a) or \ref{fig:1}(b) is equivalent to the result in Ref.~\cite{guo2022stability}.
Based on these results combined with $E_3$,
one can see an interplay among three states, that is, tripling, $s$- and $p$-wave pairings.
In Fig.~\ref{fig:1}(a) at $1/(k_{\rm F}a_s)=1.4$,
we find that the $s$-wave pairing state is stable (i.e., $|E_{2,s}|\geq |E_{2,p}|$) up to $1/(k_{\rm F}a_p)\simeq 0.25$.
Beyond this $p$-wave coupling strength, the $p$-wave Cooper pairing is stabler than the $s$-wave one (i.e., $|E_{2,s}|\geq |E_{2,p}|$).
While $E_3$ also becomes nonzero around $1/(k_{\rm F}a_s)=0.8$ and exceeds $|E_{2,s}|$ at $1/(k_{\rm F}a_s)\simeq 1$, it is larger than $E_{2,p}$
in the entire crossover region.
On the other hand, at $1/(k_{\rm F}a_s)=3$ in Fig.~\ref{fig:1}(b),
$|E_{3}|$ is larger than $|E_{2,p}|$ even in the weak-coupling $p$-wave BCS regime [i.e., $1/(k_{\rm F}a_p)\simeq -1$].
At this $s$-wave coupling strength, $|E_3|$ is larger than $|E_{2,p}|$ at stronger $p$-wave couplings.
While $|E_{2,s}|$ is larger than $E_3$ in the weak $p$-wave coupling regime,
$|E_3|$ starts to dominate over $|E_{2,s}|$ around $1/(k_{\rm F}a_p)=0.6$. 
In this regard, the Cooper triple state is stable at both strong $s$- and $p$-wave couplings.

\begin{figure}
  \includegraphics[width=0.45\textwidth]{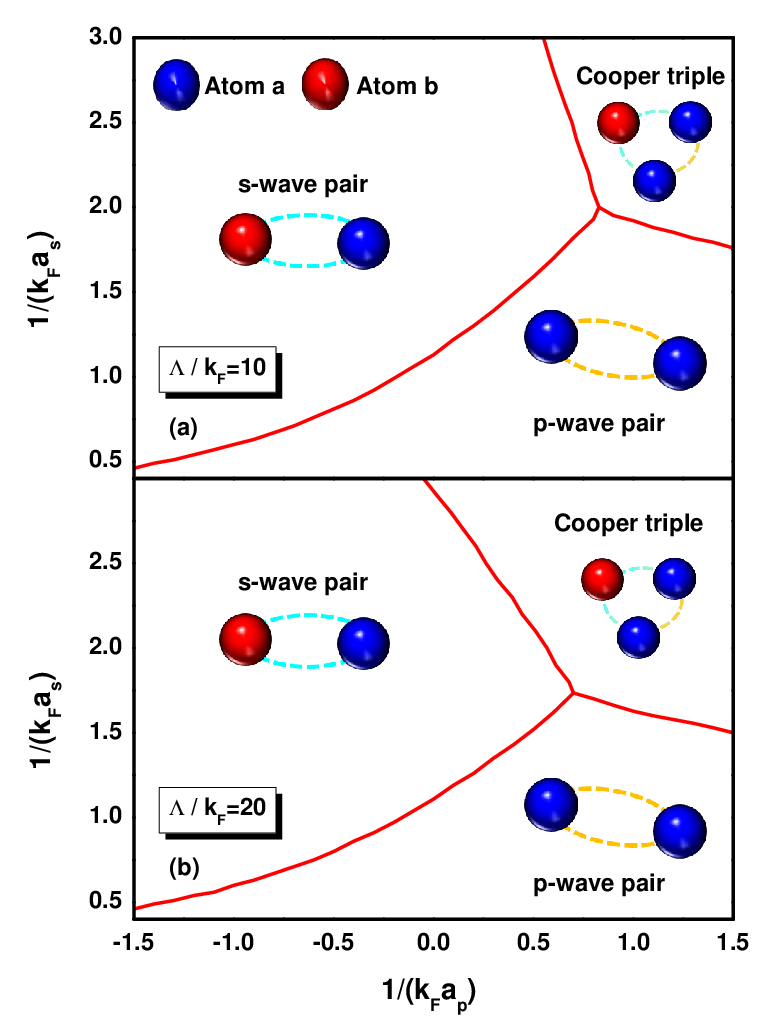}
  \caption{ 
  Phase diagram of $s$-wave pair phase ($|E_{2,s}|>|E_{2,p}|,\ |E_3|$), $p$-wave pair phase ($|E_{2,p}|>|E_{2,s}|,\ |E_3|$), and Cooper triple phase ($|E_3|>|E_{2,s}|,\ |E_{2,p}|$) in the plane of $1/(k_{\rm F}a_s)$ and $1/(k_{\rm F}a_p)$.
  The momentum cutoffs are taken as $\Lambda/k_{\rm F}=10$ and $\Lambda/k_{\rm F}=20$ in the panels (a) and (b), respectively.
  }\label{fig:2}
\end{figure}

In Fig.~\ref{fig:2},
we summarize the ground-state phase diagram of $s$- and $p$-wave Cooper pairing states, and the Cooper triple states in the present model.
The phase boundaries are determined in such a way that 
the boundary between tripling and $s$-wave pairing,
that between tripling and $p$-wave pairing,
and that between $s$- and $p$-wave pairings are given by
$E_3=E_{2,s}$,
$E_3=E_{2,p}$, and $E_{2,s}=E_{2,p}$, respectively.
While we arbitrarily employ the scattering parameters $0.5\leq 1/(k_{\rm F}a_s)\leq 3$ and $-1.5\leq 1/(k_{\rm F}a_p)\leq 1.5$ in Fig.~\ref{fig:2},
similar values were realized in a recent experimental work~\cite{Jackson2022emergent}.
Such a phase diagram captures interesting features associated with competing $s$- and $p$-wave pairings and moreover tripling.
As we showed the cutoff dependence of $E_3$ in Fig.~\ref{fig:1}(c),
the cutoff dependence of the phase diagram can be also found in Fig.~\ref{fig:2},
where we take $\Lambda=10k_{\rm F}$ and $\Lambda=20k_{\rm F}$ in Figs.~3(a) and 3(b), respectively.
The Cooper triple phase is enlarged when $\Lambda$ increases, reflecting the increase of $|E_3|$ in Fig.~\ref{fig:1}(c).
On the other hand, since the cutoff dependence of $E_{2,s}$ and $E_{2,p}$ is weaker than that of $E_3$ because of the renormalization with respect to $a_{s,p}$, the phase boundary between $s$- and $p$-wave pairings is relatively robust against the change of $\Lambda$.
While the value of $\Lambda$ is needed to compare our results with the experiments,
our phase diagram would be useful to understand the qualitative features of hybridized $s$- and $p$-wave interacting systems.

\red{
In the present framework, the $p$-wave interaction between $a$ atoms is considered.
Similarly, the Cooper triple phase, which consists of one $a$ atom and two $b$ atoms, can also be found after introducing the $p$-wave interaction between $b$ atoms.
Moreover, the possibility of the tetramer state, which we do not consider in this work, cannot be excluded. 
In particular, if both $a$-$a$ and $b$-$b$ $p$-wave interactions are present, the tetramer state may also appear, but it is out of the scope of this work.
}

\section{Summary and perspectives}\label{sec:IV}

In this paper, we have investigated competing pairing and tripling correlations in one-dimensional two-component fermions with hybridized $s$- and $p$-wave interactions.
We have solved the in-medium three-body equation derived from the variational principle based on the generalized Cooper problem. The solutions of the stable in-medium three-body cluster state (i.e., Cooper triple) have been found in this system.
Furthermore, we have shown a ground-state phase diagram consisting of $s$- and $p$-wave pairing states, and the Cooper triple states in a plane of $s$- and $p$-wave scattering lengths.
The phase diagram and the three-body ground-state energy show a cutoff dependence. 
In particular, the Cooper triple phase is enlarged by increasing the momentum cutoff.

Our work would be useful for the further investigations of superconductors and superfluids, and an understanding of the nontrivial states (such as higher-order Cooper clusters) arising from the competition and the coexistence of both $s$- and $p$-wave interactions.
Our results also suggest that low-dimensional superconductors show a Cooper triple phase due to enhanced $s$- and $p$-wave interactions by confinement or shape resonances.
Moreover, it is interesting to see how similar in-medium bound states can appear in neutron-rich matter and in lattice systems~\cite{He2019Phys.Rev.B100.201101}.

\red{
We note that our variational approach gives an approximate way to explore the ground state on top of a Fermi sea from the weak-coupling BCS pairing phase to the trimer phase in the strong-coupling limit. 
Since we treat pairing and tripling correlations by using the generalized Cooper problem, more sophisticated treatments of the many-body effects and competing orders would also be important future work.
For instance, the density-matrix renormalization group~\cite{Schollwoeck2005Rev.Mod.Phys.77.259--315} (which is efficient for the study of low-dimensional strongly correlated quantum systems) and bosonization~\cite{senechal2004introduction} (which is especially successfully applied in one-dimensional systems) can be adopted to obtain exact results to reveal many-body properties quantitatively.
}

\begin{acknowledgments}
Y.G. was supported by the RIKEN Junior Research Associate Program.
H.T. acknowledges the JSPS Grants-in-Aid for Scientific Research under Grants No.~18H05406, No.~22H01158, and No.~22K13981.
\end{acknowledgments}


\begin{thebibliography}{54}%
\makeatletter
\providecommand \@ifxundefined [1]{%
 \@ifx{#1\undefined}
}%
\providecommand \@ifnum [1]{%
 \ifnum #1\expandafter \@firstoftwo
 \else \expandafter \@secondoftwo
 \fi
}%
\providecommand \@ifx [1]{%
 \ifx #1\expandafter \@firstoftwo
 \else \expandafter \@secondoftwo
 \fi
}%
\providecommand \natexlab [1]{#1}%
\providecommand \enquote  [1]{``#1''}%
\providecommand \bibnamefont  [1]{#1}%
\providecommand \bibfnamefont [1]{#1}%
\providecommand \citenamefont [1]{#1}%
\providecommand \href@noop [0]{\@secondoftwo}%
\providecommand \href [0]{\begingroup \@sanitize@url \@href}%
\providecommand \@href[1]{\@@startlink{#1}\@@href}%
\providecommand \@@href[1]{\endgroup#1\@@endlink}%
\providecommand \@sanitize@url [0]{\catcode `\\12\catcode `\$12\catcode
  `\&12\catcode `\#12\catcode `\^12\catcode `\_12\catcode `\%12\relax}%
\providecommand \@@startlink[1]{}%
\providecommand \@@endlink[0]{}%
\providecommand \url  [0]{\begingroup\@sanitize@url \@url }%
\providecommand \@url [1]{\endgroup\@href {#1}{\urlprefix }}%
\providecommand \urlprefix  [0]{URL }%
\providecommand \Eprint [0]{\href }%
\providecommand \doibase [0]{https://doi.org/}%
\providecommand \selectlanguage [0]{\@gobble}%
\providecommand \bibinfo  [0]{\@secondoftwo}%
\providecommand \bibfield  [0]{\@secondoftwo}%
\providecommand \translation [1]{[#1]}%
\providecommand \BibitemOpen [0]{}%
\providecommand \bibitemStop [0]{}%
\providecommand \bibitemNoStop [0]{.\EOS\space}%
\providecommand \EOS [0]{\spacefactor3000\relax}%
\providecommand \BibitemShut  [1]{\csname bibitem#1\endcsname}%
\let\auto@bib@innerbib\@empty
\bibitem [{\citenamefont {Mackenzie}\ and\ \citenamefont
  {Maeno}(2003)}]{Mackenzie2003Rev.Mod.Phys.75.657--712}%
  \BibitemOpen
  \bibfield  {author} {\bibinfo {author} {\bibfnamefont {A.~P.}\ \bibnamefont
  {Mackenzie}}\ and\ \bibinfo {author} {\bibfnamefont {Y.}~\bibnamefont
  {Maeno}},\ }\bibfield  {title} {\bibinfo {title} {The superconductivity of
  sr$_2$ruo$_{4}$ and the physics of spin-triplet pairing},\ }\href
  {https://doi.org/10.1103/RevModPhys.75.657} {\bibfield  {journal} {\bibinfo
  {journal} {Rev. Mod. Phys.}\ }\textbf {\bibinfo {volume} {75}},\ \bibinfo
  {pages} {657} (\bibinfo {year} {2003})}\BibitemShut {NoStop}%
\bibitem [{\citenamefont
  {Stewart}(2011)}]{Stewart2011Rev.Mod.Phys.83.1589--1652}%
  \BibitemOpen
  \bibfield  {author} {\bibinfo {author} {\bibfnamefont {G.~R.}\ \bibnamefont
  {Stewart}},\ }\bibfield  {title} {\bibinfo {title} {Superconductivity in iron
  compounds},\ }\href {https://doi.org/10.1103/RevModPhys.83.1589} {\bibfield
  {journal} {\bibinfo  {journal} {Rev. Mod. Phys.}\ }\textbf {\bibinfo {volume}
  {83}},\ \bibinfo {pages} {1589} (\bibinfo {year} {2011})}\BibitemShut
  {NoStop}%
\bibitem [{\citenamefont {Stewart}(2017)}]{Stewart2017Adv.Phys.66.75--196}%
  \BibitemOpen
  \bibfield  {author} {\bibinfo {author} {\bibfnamefont {G.~R.}\ \bibnamefont
  {Stewart}},\ }\bibfield  {title} {\bibinfo {title} {Unconventional
  superconductivity},\ }\href {https://doi.org/10.1080/00018732.2017.1331615}
  {\bibfield  {journal} {\bibinfo  {journal} {Adv. Phys.}\ }\textbf {\bibinfo
  {volume} {66}},\ \bibinfo {pages} {75} (\bibinfo {year} {2017})}\BibitemShut
  {NoStop}%
\bibitem [{\citenamefont {Xu}\ \emph {et~al.}(2020)\citenamefont {Xu},
  \citenamefont {Zhang}, \citenamefont {Zhu},\ and\ \citenamefont
  {Guo}}]{Xu2020J.Phys.:Condens.Matter32.343003}%
  \BibitemOpen
  \bibfield  {author} {\bibinfo {author} {\bibfnamefont {X.}~\bibnamefont
  {Xu}}, \bibinfo {author} {\bibfnamefont {S.}~\bibnamefont {Zhang}}, \bibinfo
  {author} {\bibfnamefont {X.}~\bibnamefont {Zhu}},\ and\ \bibinfo {author}
  {\bibfnamefont {J.}~\bibnamefont {Guo}},\ }\bibfield  {title} {\bibinfo
  {title} {Superconductivity enhancement in {FeSe}/{SrTiO}$_3$: a review from
  the perspective of electron-phonon coupling},\ }\href
  {https://doi.org/10.1088/1361-648x/ab85f0} {\bibfield  {journal} {\bibinfo
  {journal} {J. Phys.: Condens. Matter}\ }\textbf {\bibinfo {volume} {32}},\
  \bibinfo {pages} {343003} (\bibinfo {year} {2020})}\BibitemShut {NoStop}%
\bibitem [{\citenamefont {Kanasugi}\ and\ \citenamefont
  {Yanase}(2022)}]{Kanasugi2022Comm.Phys.5.1.}%
  \BibitemOpen
  \bibfield  {author} {\bibinfo {author} {\bibfnamefont {S.}~\bibnamefont
  {Kanasugi}}\ and\ \bibinfo {author} {\bibfnamefont {Y.}~\bibnamefont
  {Yanase}},\ }\bibfield  {title} {\bibinfo {title} {Anapole superconductivity
  from $\mathcal{PT}$-symmetric mixed-parity interband pairing},\ }
   {\bibfield  {journal} {\bibinfo  {journal} {Commun. Phys.}\ }\textbf
  {\bibinfo {volume} {5}},\ \bibinfo {pages} {1} (\bibinfo {year}
  {2022})}\BibitemShut {NoStop}%
\bibitem [{\citenamefont {Takatsuka}\ and\ \citenamefont
  {Tamagaki}(1993)}]{Takatsuka1993PTP.112.27}%
  \BibitemOpen
  \bibfield  {author} {\bibinfo {author} {\bibfnamefont {T.}~\bibnamefont
  {Takatsuka}}\ and\ \bibinfo {author} {\bibfnamefont {R.}~\bibnamefont
  {Tamagaki}},\ }\bibfield  {title} {\bibinfo {title} {{Superfluidity in
  Neutron Star Matter and Symmetric Nuclear Matter}},\ }\href
  {https://doi.org/10.1143/PTP.112.27} {\bibfield  {journal} {\bibinfo
  {journal} {Prog. Theor. Phys. Suppl.}\ }\textbf {\bibinfo {volume} {112}},\
  \bibinfo {pages} {27} (\bibinfo {year} {1993})}\BibitemShut {NoStop}%
\bibitem [{\citenamefont {Yasui}\ \emph {et~al.}(2020)\citenamefont {Yasui},
  \citenamefont {Inotani},\ and\ \citenamefont
  {Nitta}}]{Yasui2020PhysRevC.101.055806}%
  \BibitemOpen
  \bibfield  {author} {\bibinfo {author} {\bibfnamefont {S.}~\bibnamefont
  {Yasui}}, \bibinfo {author} {\bibfnamefont {D.}~\bibnamefont {Inotani}},\
  and\ \bibinfo {author} {\bibfnamefont {M.}~\bibnamefont {Nitta}},\ }\bibfield
   {title} {\bibinfo {title} {Coexistence phase of $^{1}$s$_{0}$ and
  $^{3}$p$_{2}$ superfluids in neutron stars},\ }\href
  {https://doi.org/10.1103/PhysRevC.101.055806} {\bibfield  {journal} {\bibinfo
   {journal} {Phys. Rev. C}\ }\textbf {\bibinfo {volume} {101}},\ \bibinfo
  {pages} {055806} (\bibinfo {year} {2020})}\BibitemShut {NoStop}%
\bibitem [{\citenamefont {Hammer}\ \emph {et~al.}(2017)\citenamefont {Hammer},
  \citenamefont {Ji},\ and\ \citenamefont {Phillips}}]{Hammer2017}%
  \BibitemOpen
  \bibfield  {author} {\bibinfo {author} {\bibfnamefont {H.-W.}\ \bibnamefont
  {Hammer}}, \bibinfo {author} {\bibfnamefont {C.}~\bibnamefont {Ji}},\ and\
  \bibinfo {author} {\bibfnamefont {D.~R.}\ \bibnamefont {Phillips}},\
  }\bibfield  {title} {\bibinfo {title} {Effective field theory description of
  halo nuclei},\ }\href {https://doi.org/10.1088/1361-6471/aa83db} {\bibfield
  {journal} {\bibinfo  {journal} {Journal of Physics G: Nuclear and Particle
  Physics}\ }\textbf {\bibinfo {volume} {44}},\ \bibinfo {pages} {103002}
  (\bibinfo {year} {2017})}\BibitemShut {NoStop}%
\bibitem [{\citenamefont {Marqu{\'e}s}\ and\ \citenamefont
  {Carbonell}(2021)}]{marques2021quest}%
  \BibitemOpen
  \bibfield  {author} {\bibinfo {author} {\bibfnamefont {F.~M.}\ \bibnamefont
  {Marqu{\'e}s}}\ and\ \bibinfo {author} {\bibfnamefont {J.}~\bibnamefont
  {Carbonell}},\ }\bibfield  {title} {\bibinfo {title} {The quest for light
  multineutron systems},\ } {\bibfield  {journal} {\bibinfo
  {journal} {The European Physical Journal A}\ }\textbf {\bibinfo {volume}
  {57}},\ \bibinfo {pages} {1} (\bibinfo {year} {2021})}\BibitemShut {NoStop}%
\bibitem [{\citenamefont {Duer}\ \emph {et~al.}(2022)\citenamefont {Duer},
  \citenamefont {Aumann}, \citenamefont {Gernh{\"a}user}, \citenamefont
  {Panin}, \citenamefont {Paschalis}, \citenamefont {Rossi}, \citenamefont
  {Achouri}, \citenamefont {Ahn}, \citenamefont {Baba}, \citenamefont
  {Bertulani} \emph {et~al.}}]{duer2022observation}%
  \BibitemOpen
  \bibfield  {author} {\bibinfo {author} {\bibfnamefont {M.}~\bibnamefont
  {Duer}}, \bibinfo {author} {\bibfnamefont {T.}~\bibnamefont {Aumann}},
  \bibinfo {author} {\bibfnamefont {R.}~\bibnamefont {Gernh{\"a}user}},
  \bibinfo {author} {\bibfnamefont {V.}~\bibnamefont {Panin}}, \bibinfo
  {author} {\bibfnamefont {S.}~\bibnamefont {Paschalis}}, \bibinfo {author}
  {\bibfnamefont {D.}~\bibnamefont {Rossi}}, \bibinfo {author} {\bibfnamefont
  {N.}~\bibnamefont {Achouri}}, \bibinfo {author} {\bibfnamefont
  {D.}~\bibnamefont {Ahn}}, \bibinfo {author} {\bibfnamefont {H.}~\bibnamefont
  {Baba}}, \bibinfo {author} {\bibfnamefont {C.}~\bibnamefont {Bertulani}},
  \emph {et~al.},\ }\bibfield  {title} {\bibinfo {title} {Observation of a
  correlated free four-neutron system},\ }{\bibfield  {journal}
  {\bibinfo  {journal} {Nature}\ }\textbf {\bibinfo {volume} {606}},\ \bibinfo
  {pages} {678} (\bibinfo {year} {2022})}\BibitemShut {NoStop}%
\bibitem [{\citenamefont {Chin}\ \emph {et~al.}(2010)\citenamefont {Chin},
  \citenamefont {Grimm}, \citenamefont {Julienne},\ and\ \citenamefont
  {Tiesinga}}]{Chin2010Rev.Mod.Phys.82.1225--1286}%
  \BibitemOpen
  \bibfield  {author} {\bibinfo {author} {\bibfnamefont {C.}~\bibnamefont
  {Chin}}, \bibinfo {author} {\bibfnamefont {R.}~\bibnamefont {Grimm}},
  \bibinfo {author} {\bibfnamefont {P.}~\bibnamefont {Julienne}},\ and\
  \bibinfo {author} {\bibfnamefont {E.}~\bibnamefont {Tiesinga}},\ }\bibfield
  {title} {\bibinfo {title} {Feshbach resonances in ultracold gases},\ }\href
  {https://doi.org/10.1103/RevModPhys.82.1225} {\bibfield  {journal} {\bibinfo
  {journal} {Rev. Mod. Phys.}\ }\textbf {\bibinfo {volume} {82}},\ \bibinfo
  {pages} {1225} (\bibinfo {year} {2010})}\BibitemShut {NoStop}%
\bibitem [{\citenamefont {Randeria}\ and\ \citenamefont
  {Taylor}(2014)}]{doi:10.1146/annurev-conmatphys-031113-133829}%
  \BibitemOpen
  \bibfield  {author} {\bibinfo {author} {\bibfnamefont {M.}~\bibnamefont
  {Randeria}}\ and\ \bibinfo {author} {\bibfnamefont {E.}~\bibnamefont
  {Taylor}},\ }\bibfield  {title} {\bibinfo {title} {Crossover from
  bardeen-cooper-schrieffer to bose-einstein condensation and the unitary fermi
  gas},\ }\href {https://doi.org/10.1146/annurev-conmatphys-031113-133829}
  {\bibfield  {journal} {\bibinfo  {journal} {Annual Review of Condensed Matter
  Physics}\ }\textbf {\bibinfo {volume} {5}},\ \bibinfo {pages} {209} (\bibinfo
  {year} {2014})}\BibitemShut {NoStop}%
\bibitem [{\citenamefont {Strinati}\ \emph {et~al.}(2018)\citenamefont
  {Strinati}, \citenamefont {Pieri}, \citenamefont {Röpke}, \citenamefont
  {Schuck},\ and\ \citenamefont {Urban}}]{Strinati2018Phys.Rep.738.1--76}%
  \BibitemOpen
  \bibfield  {author} {\bibinfo {author} {\bibfnamefont {G.~C.}\ \bibnamefont
  {Strinati}}, \bibinfo {author} {\bibfnamefont {P.}~\bibnamefont {Pieri}},
  \bibinfo {author} {\bibfnamefont {G.}~\bibnamefont {Röpke}}, \bibinfo
  {author} {\bibfnamefont {P.}~\bibnamefont {Schuck}},\ and\ \bibinfo {author}
  {\bibfnamefont {M.}~\bibnamefont {Urban}},\ }\bibfield  {title} {\bibinfo
  {title} {The bcs–bec crossover: From ultra-cold fermi gases to nuclear
  systems},\ }\href
  {https://doi.org/https://doi.org/10.1016/j.physrep.2018.02.004} {\bibfield
  {journal} {\bibinfo  {journal} {Phys. Rep.}\ }\textbf {\bibinfo {volume}
  {738}},\ \bibinfo {pages} {1} (\bibinfo {year} {2018})}\BibitemShut {NoStop}%
\bibitem [{\citenamefont {Ohashi}\ \emph {et~al.}(2020)\citenamefont {Ohashi},
  \citenamefont {Tajima},\ and\ \citenamefont {{van
  Wyk}}}]{Ohashi2020Prog.Part.Nucl.Phys.111.103739}%
  \BibitemOpen
  \bibfield  {author} {\bibinfo {author} {\bibfnamefont {Y.}~\bibnamefont
  {Ohashi}}, \bibinfo {author} {\bibfnamefont {H.}~\bibnamefont {Tajima}},\
  and\ \bibinfo {author} {\bibfnamefont {P.}~\bibnamefont {{van Wyk}}},\
  }\bibfield  {title} {\bibinfo {title} {Bcs–bec crossover in cold atomic and
  in nuclear systems},\ }\href
  {https://doi.org/https://doi.org/10.1016/j.ppnp.2019.103739} {\bibfield
  {journal} {\bibinfo  {journal} {Prog. Part. Nucl. Phys.}\ }\textbf {\bibinfo
  {volume} {111}},\ \bibinfo {pages} {103739} (\bibinfo {year}
  {2020})}\BibitemShut {NoStop}%
\bibitem [{\citenamefont {Ticknor}\ \emph {et~al.}(2004)\citenamefont
  {Ticknor}, \citenamefont {Regal}, \citenamefont {Jin},\ and\ \citenamefont
  {Bohn}}]{Ticknor2004Phys.Rev.A69.042712}%
  \BibitemOpen
  \bibfield  {author} {\bibinfo {author} {\bibfnamefont {C.}~\bibnamefont
  {Ticknor}}, \bibinfo {author} {\bibfnamefont {C.~A.}\ \bibnamefont {Regal}},
  \bibinfo {author} {\bibfnamefont {D.~S.}\ \bibnamefont {Jin}},\ and\ \bibinfo
  {author} {\bibfnamefont {J.~L.}\ \bibnamefont {Bohn}},\ }\bibfield  {title}
  {\bibinfo {title} {Multiplet structure of feshbach resonances in nonzero
  partial waves},\ }\href {https://doi.org/10.1103/PhysRevA.69.042712}
  {\bibfield  {journal} {\bibinfo  {journal} {Phys. Rev. A}\ }\textbf {\bibinfo
  {volume} {69}},\ \bibinfo {pages} {042712} (\bibinfo {year}
  {2004})}\BibitemShut {NoStop}%
\bibitem [{\citenamefont {Gurarie}\ \emph {et~al.}(2005)\citenamefont
  {Gurarie}, \citenamefont {Radzihovsky},\ and\ \citenamefont
  {Andreev}}]{Gurarie2005Phys.Rev.Lett.94.230403}%
  \BibitemOpen
  \bibfield  {author} {\bibinfo {author} {\bibfnamefont {V.}~\bibnamefont
  {Gurarie}}, \bibinfo {author} {\bibfnamefont {L.}~\bibnamefont
  {Radzihovsky}},\ and\ \bibinfo {author} {\bibfnamefont {A.~V.}\ \bibnamefont
  {Andreev}},\ }\bibfield  {title} {\bibinfo {title} {Quantum phase transitions
  across a $p$-wave feshbach resonance},\ }\href
  {https://doi.org/10.1103/PhysRevLett.94.230403} {\bibfield  {journal}
  {\bibinfo  {journal} {Phys. Rev. Lett.}\ }\textbf {\bibinfo {volume} {94}},\
  \bibinfo {pages} {230403} (\bibinfo {year} {2005})}\BibitemShut {NoStop}%
\bibitem [{\citenamefont {Schunck}\ \emph {et~al.}(2005)\citenamefont
  {Schunck}, \citenamefont {Zwierlein}, \citenamefont {Stan}, \citenamefont
  {Raupach}, \citenamefont {Ketterle}, \citenamefont {Simoni}, \citenamefont
  {Tiesinga}, \citenamefont {Williams},\ and\ \citenamefont
  {Julienne}}]{Schunck2005Phys.Rev.A71.045601}%
  \BibitemOpen
  \bibfield  {author} {\bibinfo {author} {\bibfnamefont {C.~H.}\ \bibnamefont
  {Schunck}}, \bibinfo {author} {\bibfnamefont {M.~W.}\ \bibnamefont
  {Zwierlein}}, \bibinfo {author} {\bibfnamefont {C.~A.}\ \bibnamefont {Stan}},
  \bibinfo {author} {\bibfnamefont {S.~M.~F.}\ \bibnamefont {Raupach}},
  \bibinfo {author} {\bibfnamefont {W.}~\bibnamefont {Ketterle}}, \bibinfo
  {author} {\bibfnamefont {A.}~\bibnamefont {Simoni}}, \bibinfo {author}
  {\bibfnamefont {E.}~\bibnamefont {Tiesinga}}, \bibinfo {author}
  {\bibfnamefont {C.~J.}\ \bibnamefont {Williams}},\ and\ \bibinfo {author}
  {\bibfnamefont {P.~S.}\ \bibnamefont {Julienne}},\ }\bibfield  {title}
  {\bibinfo {title} {Feshbach resonances in fermionic $^{6}\mathrm{Li}$},\
  }\href {https://doi.org/10.1103/PhysRevA.71.045601} {\bibfield  {journal}
  {\bibinfo  {journal} {Phys. Rev. A}\ }\textbf {\bibinfo {volume} {71}},\
  \bibinfo {pages} {045601} (\bibinfo {year} {2005})}\BibitemShut {NoStop}%
\bibitem [{\citenamefont {Inada}\ \emph {et~al.}(2008)\citenamefont {Inada},
  \citenamefont {Horikoshi}, \citenamefont {Nakajima}, \citenamefont
  {Kuwata-Gonokami}, \citenamefont {Ueda},\ and\ \citenamefont
  {Mukaiyama}}]{Inada2008PhysRevLett.101.100401}%
  \BibitemOpen
  \bibfield  {author} {\bibinfo {author} {\bibfnamefont {Y.}~\bibnamefont
  {Inada}}, \bibinfo {author} {\bibfnamefont {M.}~\bibnamefont {Horikoshi}},
  \bibinfo {author} {\bibfnamefont {S.}~\bibnamefont {Nakajima}}, \bibinfo
  {author} {\bibfnamefont {M.}~\bibnamefont {Kuwata-Gonokami}}, \bibinfo
  {author} {\bibfnamefont {M.}~\bibnamefont {Ueda}},\ and\ \bibinfo {author}
  {\bibfnamefont {T.}~\bibnamefont {Mukaiyama}},\ }\bibfield  {title} {\bibinfo
  {title} {Collisional properties of $p$-wave feshbach molecules},\ }\href
  {https://doi.org/10.1103/PhysRevLett.101.100401} {\bibfield  {journal}
  {\bibinfo  {journal} {Phys. Rev. Lett.}\ }\textbf {\bibinfo {volume} {101}},\
  \bibinfo {pages} {100401} (\bibinfo {year} {2008})}\BibitemShut {NoStop}%
\bibitem [{\citenamefont {Nakasuji}\ \emph {et~al.}(2013)\citenamefont
  {Nakasuji}, \citenamefont {Yoshida},\ and\ \citenamefont
  {Mukaiyama}}]{Nakasuji2013PhysRevA.88.012710}%
  \BibitemOpen
  \bibfield  {author} {\bibinfo {author} {\bibfnamefont {T.}~\bibnamefont
  {Nakasuji}}, \bibinfo {author} {\bibfnamefont {J.}~\bibnamefont {Yoshida}},\
  and\ \bibinfo {author} {\bibfnamefont {T.}~\bibnamefont {Mukaiyama}},\
  }\bibfield  {title} {\bibinfo {title} {Experimental determination of $p$-wave
  scattering parameters in ultracold ${}^{6}$li atoms},\ }\href
  {https://doi.org/10.1103/PhysRevA.88.012710} {\bibfield  {journal} {\bibinfo
  {journal} {Phys. Rev. A}\ }\textbf {\bibinfo {volume} {88}},\ \bibinfo
  {pages} {012710} (\bibinfo {year} {2013})}\BibitemShut {NoStop}%
\bibitem [{\citenamefont {Regal}\ \emph {et~al.}(2003)\citenamefont {Regal},
  \citenamefont {Ticknor}, \citenamefont {Bohn},\ and\ \citenamefont
  {Jin}}]{Regal2003PhysRevLett.90.053201}%
  \BibitemOpen
  \bibfield  {author} {\bibinfo {author} {\bibfnamefont {C.~A.}\ \bibnamefont
  {Regal}}, \bibinfo {author} {\bibfnamefont {C.}~\bibnamefont {Ticknor}},
  \bibinfo {author} {\bibfnamefont {J.~L.}\ \bibnamefont {Bohn}},\ and\
  \bibinfo {author} {\bibfnamefont {D.~S.}\ \bibnamefont {Jin}},\ }\bibfield
  {title} {\bibinfo {title} {Tuning $p$-wave interactions in an ultracold fermi
  gas of atoms},\ }\href {https://doi.org/10.1103/PhysRevLett.90.053201}
  {\bibfield  {journal} {\bibinfo  {journal} {Phys. Rev. Lett.}\ }\textbf
  {\bibinfo {volume} {90}},\ \bibinfo {pages} {053201} (\bibinfo {year}
  {2003})}\BibitemShut {NoStop}%
\bibitem [{\citenamefont {Zhou}\ \emph {et~al.}(2017)\citenamefont {Zhou},
  \citenamefont {Yi},\ and\ \citenamefont {Cui}}]{Zhou2017ScienceChina.60.12}%
  \BibitemOpen
  \bibfield  {author} {\bibinfo {author} {\bibfnamefont {L.}~\bibnamefont
  {Zhou}}, \bibinfo {author} {\bibfnamefont {W.}~\bibnamefont {Yi}},\ and\
  \bibinfo {author} {\bibfnamefont {X.}~\bibnamefont {Cui}},\ }\bibfield
  {title} {\bibinfo {title} {Fermion superfluid with hybridized s-and p-wave
  pairings},\ }{\bibfield  {journal} {\bibinfo  {journal}
  {SCIENCE CHINA Physics, Mechanics \& Astronomy}\ }\textbf {\bibinfo {volume}
  {60}},\ \bibinfo {pages} {1} (\bibinfo {year} {2017})}\BibitemShut {NoStop}%
\bibitem [{\citenamefont {Naidon}\ \emph {et~al.}(2022)\citenamefont {Naidon},
  \citenamefont {Pricoupenko},\ and\ \citenamefont
  {Schmickler}}]{10.21468/SciPostPhys.12.6.185}%
  \BibitemOpen
  \bibfield  {author} {\bibinfo {author} {\bibfnamefont {P.}~\bibnamefont
  {Naidon}}, \bibinfo {author} {\bibfnamefont {L.}~\bibnamefont
  {Pricoupenko}},\ and\ \bibinfo {author} {\bibfnamefont {C.}~\bibnamefont
  {Schmickler}},\ }\bibfield  {title} {\bibinfo {title} {{Shallow trimers of
  two identical fermions and one particle in resonant regimes}},\ }\href
  {https://doi.org/10.21468/SciPostPhys.12.6.185} {\bibfield  {journal}
  {\bibinfo  {journal} {SciPost Phys.}\ }\textbf {\bibinfo {volume} {12}},\
  \bibinfo {pages} {185} (\bibinfo {year} {2022})}\BibitemShut {NoStop}%
\bibitem [{\citenamefont {Jackson}\ \emph {et~al.}(2022)\citenamefont
  {Jackson}, \citenamefont {Dale}, \citenamefont {Maki}, \citenamefont {Xie},
  \citenamefont {Olsen}, \citenamefont {Ahmed-Braun}, \citenamefont {Zhang},\
  and\ \citenamefont {Thywissen}}]{Jackson2022emergent}%
  \BibitemOpen
  \bibfield  {author} {\bibinfo {author} {\bibfnamefont {K.~G.}\ \bibnamefont
  {Jackson}}, \bibinfo {author} {\bibfnamefont {C.~J.}\ \bibnamefont {Dale}},
  \bibinfo {author} {\bibfnamefont {J.}~\bibnamefont {Maki}}, \bibinfo {author}
  {\bibfnamefont {K.~G.}\ \bibnamefont {Xie}}, \bibinfo {author} {\bibfnamefont
  {B.~A.}\ \bibnamefont {Olsen}}, \bibinfo {author} {\bibfnamefont {D.~J.}\
  \bibnamefont {Ahmed-Braun}}, \bibinfo {author} {\bibfnamefont
  {S.}~\bibnamefont {Zhang}},\ and\ \bibinfo {author} {\bibfnamefont {J.~H.}\
  \bibnamefont {Thywissen}},\ }\bibfield  {title} {\bibinfo {title} {Emergent
  s-wave interactions between identical fermions in quasi-one-dimensional
  geometries},\ }{\bibfield  {journal} {\bibinfo  {journal}
  {arXiv preprint arXiv:2206.10415}\ } (\bibinfo {year} {2022})}\BibitemShut
  {NoStop}%
\bibitem [{\citenamefont {Niemann}\ and\ \citenamefont
  {Hammer}(2012)}]{Niemann2012Phys.Rev.A86.013628}%
  \BibitemOpen
  \bibfield  {author} {\bibinfo {author} {\bibfnamefont {P.}~\bibnamefont
  {Niemann}}\ and\ \bibinfo {author} {\bibfnamefont {H.-W.}\ \bibnamefont
  {Hammer}},\ }\bibfield  {title} {\bibinfo {title} {Pauli-blocking effects and
  cooper triples in three-component fermi gases},\ }\href
  {https://doi.org/10.1103/PhysRevA.86.013628} {\bibfield  {journal} {\bibinfo
  {journal} {Phys. Rev. A}\ }\textbf {\bibinfo {volume} {86}},\ \bibinfo
  {pages} {013628} (\bibinfo {year} {2012})}\BibitemShut {NoStop}%
\bibitem [{\citenamefont {Kirk}\ and\ \citenamefont
  {Parish}(2017)}]{Kirk2017Phys.Rev.A96.053614}%
  \BibitemOpen
  \bibfield  {author} {\bibinfo {author} {\bibfnamefont {T.}~\bibnamefont
  {Kirk}}\ and\ \bibinfo {author} {\bibfnamefont {M.~M.}\ \bibnamefont
  {Parish}},\ }\bibfield  {title} {\bibinfo {title} {Three-body correlations in
  a two-dimensional su(3) fermi gas},\ }\href
  {https://doi.org/10.1103/PhysRevA.96.053614} {\bibfield  {journal} {\bibinfo
  {journal} {Phys. Rev. A}\ }\textbf {\bibinfo {volume} {96}},\ \bibinfo
  {pages} {053614} (\bibinfo {year} {2017})}\BibitemShut {NoStop}%
\bibitem [{\citenamefont {Akagami}\ \emph {et~al.}(2021)\citenamefont
  {Akagami}, \citenamefont {Tajima},\ and\ \citenamefont
  {Iida}}]{Akagami2021Phys.Rev.A104.L041302}%
  \BibitemOpen
  \bibfield  {author} {\bibinfo {author} {\bibfnamefont {S.}~\bibnamefont
  {Akagami}}, \bibinfo {author} {\bibfnamefont {H.}~\bibnamefont {Tajima}},\
  and\ \bibinfo {author} {\bibfnamefont {K.}~\bibnamefont {Iida}},\ }\bibfield
  {title} {\bibinfo {title} {Condensation of cooper triples},\ }\href
  {https://doi.org/10.1103/PhysRevA.104.L041302} {\bibfield  {journal}
  {\bibinfo  {journal} {Phys. Rev. A}\ }\textbf {\bibinfo {volume} {104}},\
  \bibinfo {pages} {L041302} (\bibinfo {year} {2021})}\BibitemShut {NoStop}%
\bibitem [{\citenamefont {Tajima}\ \emph
  {et~al.}(2021{\natexlab{a}})\citenamefont {Tajima}, \citenamefont {Tsutsui},
  \citenamefont {Doi},\ and\ \citenamefont
  {Iida}}]{Tajima2021Phys.Rev.A104.053328}%
  \BibitemOpen
  \bibfield  {author} {\bibinfo {author} {\bibfnamefont {H.}~\bibnamefont
  {Tajima}}, \bibinfo {author} {\bibfnamefont {S.}~\bibnamefont {Tsutsui}},
  \bibinfo {author} {\bibfnamefont {T.~M.}\ \bibnamefont {Doi}},\ and\ \bibinfo
  {author} {\bibfnamefont {K.}~\bibnamefont {Iida}},\ }\bibfield  {title}
  {\bibinfo {title} {Three-body crossover from a cooper triple to a bound
  trimer state in three-component fermi gases near a triatomic resonance},\
  }\href {https://doi.org/10.1103/PhysRevA.104.053328} {\bibfield  {journal}
  {\bibinfo  {journal} {Phys. Rev. A}\ }\textbf {\bibinfo {volume} {104}},\
  \bibinfo {pages} {053328} (\bibinfo {year} {2021}{\natexlab{a}})}\BibitemShut
  {NoStop}%
\bibitem [{\citenamefont {Tajima}\ \emph {et~al.}(2022)\citenamefont {Tajima},
  \citenamefont {Tsutsui}, \citenamefont {Doi},\ and\ \citenamefont
  {Iida}}]{Tajima2022Phys.Rev.Research4.L012021}%
  \BibitemOpen
  \bibfield  {author} {\bibinfo {author} {\bibfnamefont {H.}~\bibnamefont
  {Tajima}}, \bibinfo {author} {\bibfnamefont {S.}~\bibnamefont {Tsutsui}},
  \bibinfo {author} {\bibfnamefont {T.~M.}\ \bibnamefont {Doi}},\ and\ \bibinfo
  {author} {\bibfnamefont {K.}~\bibnamefont {Iida}},\ }\bibfield  {title}
  {\bibinfo {title} {Cooper triples in attractive three-component fermions:
  Implication for hadron-quark crossover},\ }\href
  {https://doi.org/10.1103/PhysRevResearch.4.L012021} {\bibfield  {journal}
  {\bibinfo  {journal} {Phys. Rev. Research}\ }\textbf {\bibinfo {volume}
  {4}},\ \bibinfo {pages} {L012021} (\bibinfo {year} {2022})}\BibitemShut
  {NoStop}%
\bibitem [{\citenamefont {R\"opke}\ \emph {et~al.}(1998)\citenamefont
  {R\"opke}, \citenamefont {Schnell}, \citenamefont {Schuck},\ and\
  \citenamefont {Nozi\`eres}}]{Roepke1998Phys.Rev.Lett.80.3177--3180}%
  \BibitemOpen
  \bibfield  {author} {\bibinfo {author} {\bibfnamefont {G.}~\bibnamefont
  {R\"opke}}, \bibinfo {author} {\bibfnamefont {A.}~\bibnamefont {Schnell}},
  \bibinfo {author} {\bibfnamefont {P.}~\bibnamefont {Schuck}},\ and\ \bibinfo
  {author} {\bibfnamefont {P.}~\bibnamefont {Nozi\`eres}},\ }\bibfield  {title}
  {\bibinfo {title} {Four-particle condensate in strongly coupled fermion
  systems},\ }\href {https://doi.org/10.1103/PhysRevLett.80.3177} {\bibfield
  {journal} {\bibinfo  {journal} {Phys. Rev. Lett.}\ }\textbf {\bibinfo
  {volume} {80}},\ \bibinfo {pages} {3177} (\bibinfo {year}
  {1998})}\BibitemShut {NoStop}%
\bibitem [{\citenamefont {Sandulescu}\ \emph {et~al.}(2012)\citenamefont
  {Sandulescu}, \citenamefont {Negrea}, \citenamefont {Dukelsky},\ and\
  \citenamefont {Johnson}}]{Sandulescu2012Phys.Rev.C85.061303}%
  \BibitemOpen
  \bibfield  {author} {\bibinfo {author} {\bibfnamefont {N.}~\bibnamefont
  {Sandulescu}}, \bibinfo {author} {\bibfnamefont {D.}~\bibnamefont {Negrea}},
  \bibinfo {author} {\bibfnamefont {J.}~\bibnamefont {Dukelsky}},\ and\
  \bibinfo {author} {\bibfnamefont {C.~W.}\ \bibnamefont {Johnson}},\
  }\bibfield  {title} {\bibinfo {title} {Quartet condensation and isovector
  pairing correlations in $n=z$ nuclei},\ }\href
  {https://doi.org/10.1103/PhysRevC.85.061303} {\bibfield  {journal} {\bibinfo
  {journal} {Phys. Rev. C}\ }\textbf {\bibinfo {volume} {85}},\ \bibinfo
  {pages} {061303} (\bibinfo {year} {2012})}\BibitemShut {NoStop}%
\bibitem [{\citenamefont {Baran}\ and\ \citenamefont
  {Delion}(2020)}]{Baran2020Phys.Lett.B805.135462}%
  \BibitemOpen
  \bibfield  {author} {\bibinfo {author} {\bibfnamefont {V.}~\bibnamefont
  {Baran}}\ and\ \bibinfo {author} {\bibfnamefont {D.}~\bibnamefont {Delion}},\
  }\bibfield  {title} {\bibinfo {title} {A quartet bcs-like theory},\ }\href
  {https://doi.org/https://doi.org/10.1016/j.physletb.2020.135462} {\bibfield
  {journal} {\bibinfo  {journal} {Phys. Lett. B}\ }\textbf {\bibinfo {volume}
  {805}},\ \bibinfo {pages} {135462} (\bibinfo {year} {2020})}\BibitemShut
  {NoStop}%
\bibitem [{\citenamefont {Guo}\ \emph {et~al.}(2022{\natexlab{a}})\citenamefont
  {Guo}, \citenamefont {Tajima},\ and\ \citenamefont
  {Liang}}]{Guo2022Phys.Rev.C105.024317}%
  \BibitemOpen
  \bibfield  {author} {\bibinfo {author} {\bibfnamefont {Y.}~\bibnamefont
  {Guo}}, \bibinfo {author} {\bibfnamefont {H.}~\bibnamefont {Tajima}},\ and\
  \bibinfo {author} {\bibfnamefont {H.}~\bibnamefont {Liang}},\ }\bibfield
  {title} {\bibinfo {title} {Cooper quartet correlations in infinite symmetric
  nuclear matter},\ }\href {https://doi.org/10.1103/PhysRevC.105.024317}
  {\bibfield  {journal} {\bibinfo  {journal} {Phys. Rev. C}\ }\textbf {\bibinfo
  {volume} {105}},\ \bibinfo {pages} {024317} (\bibinfo {year}
  {2022}{\natexlab{a}})}\BibitemShut {NoStop}%
\bibitem [{\citenamefont {Guo}\ \emph {et~al.}(2022{\natexlab{b}})\citenamefont
  {Guo}, \citenamefont {Tajima},\ and\ \citenamefont
  {Liang}}]{Guo2022Phys.Rev.Research4.023152}%
  \BibitemOpen
  \bibfield  {author} {\bibinfo {author} {\bibfnamefont {Y.}~\bibnamefont
  {Guo}}, \bibinfo {author} {\bibfnamefont {H.}~\bibnamefont {Tajima}},\ and\
  \bibinfo {author} {\bibfnamefont {H.}~\bibnamefont {Liang}},\ }\bibfield
  {title} {\bibinfo {title} {Biexciton-like quartet condensates in an
  electron-hole liquid},\ }\href
  {https://doi.org/10.1103/PhysRevResearch.4.023152} {\bibfield  {journal}
  {\bibinfo  {journal} {Phys. Rev. Research}\ }\textbf {\bibinfo {volume}
  {4}},\ \bibinfo {pages} {023152} (\bibinfo {year}
  {2022}{\natexlab{b}})}\BibitemShut {NoStop}%
\bibitem [{\citenamefont {Naidon}\ and\ \citenamefont
  {Endo}(2017)}]{Naidon2017Rep.Prog.Phys.80.056001}%
  \BibitemOpen
  \bibfield  {author} {\bibinfo {author} {\bibfnamefont {P.}~\bibnamefont
  {Naidon}}\ and\ \bibinfo {author} {\bibfnamefont {S.}~\bibnamefont {Endo}},\
  }\bibfield  {title} {\bibinfo {title} {Efimov physics: a review},\
  } {\bibfield  {journal} {\bibinfo  {journal} {Rep. Prog. Phys.}\
  }\textbf {\bibinfo {volume} {80}},\ \bibinfo {pages} {056001} (\bibinfo
  {year} {2017})}\BibitemShut {NoStop}%
\bibitem [{\citenamefont {Bloch}\ \emph {et~al.}(2008)\citenamefont {Bloch},
  \citenamefont {Dalibard},\ and\ \citenamefont
  {Zwerger}}]{Bloch2008Rev.Mod.Phys.80.885--964}%
  \BibitemOpen
  \bibfield  {author} {\bibinfo {author} {\bibfnamefont {I.}~\bibnamefont
  {Bloch}}, \bibinfo {author} {\bibfnamefont {J.}~\bibnamefont {Dalibard}},\
  and\ \bibinfo {author} {\bibfnamefont {W.}~\bibnamefont {Zwerger}},\
  }\bibfield  {title} {\bibinfo {title} {Many-body physics with ultracold
  gases},\ }\href {https://doi.org/10.1103/RevModPhys.80.885} {\bibfield
  {journal} {\bibinfo  {journal} {Rev. Mod. Phys.}\ }\textbf {\bibinfo {volume}
  {80}},\ \bibinfo {pages} {885} (\bibinfo {year} {2008})}\BibitemShut
  {NoStop}%
\bibitem [{\citenamefont {Olshanii}(1998)}]{Olshanii1998PhysRevLett.81.938}%
  \BibitemOpen
  \bibfield  {author} {\bibinfo {author} {\bibfnamefont {M.}~\bibnamefont
  {Olshanii}},\ }\bibfield  {title} {\bibinfo {title} {Atomic scattering in the
  presence of an external confinement and a gas of impenetrable bosons},\
  }\href {https://doi.org/10.1103/PhysRevLett.81.938} {\bibfield  {journal}
  {\bibinfo  {journal} {Phys. Rev. Lett.}\ }\textbf {\bibinfo {volume} {81}},\
  \bibinfo {pages} {938} (\bibinfo {year} {1998})}\BibitemShut {NoStop}%
\bibitem [{\citenamefont {Bergeman}\ \emph {et~al.}(2003)\citenamefont
  {Bergeman}, \citenamefont {Moore},\ and\ \citenamefont
  {Olshanii}}]{Bergeman2003PhysRevLett.91.163201}%
  \BibitemOpen
  \bibfield  {author} {\bibinfo {author} {\bibfnamefont {T.}~\bibnamefont
  {Bergeman}}, \bibinfo {author} {\bibfnamefont {M.~G.}\ \bibnamefont
  {Moore}},\ and\ \bibinfo {author} {\bibfnamefont {M.}~\bibnamefont
  {Olshanii}},\ }\bibfield  {title} {\bibinfo {title} {Atom-atom scattering
  under cylindrical harmonic confinement: Numerical and analytic studies of the
  confinement induced resonance},\ }\href
  {https://doi.org/10.1103/PhysRevLett.91.163201} {\bibfield  {journal}
  {\bibinfo  {journal} {Phys. Rev. Lett.}\ }\textbf {\bibinfo {volume} {91}},\
  \bibinfo {pages} {163201} (\bibinfo {year} {2003})}\BibitemShut {NoStop}%
\bibitem [{\citenamefont {Moritz}\ \emph {et~al.}(2005)\citenamefont {Moritz},
  \citenamefont {St\"oferle}, \citenamefont {G\"unter}, \citenamefont
  {K\"ohl},\ and\ \citenamefont {Esslinger}}]{Moritz2005PhysRevLett.94.210401}%
  \BibitemOpen
  \bibfield  {author} {\bibinfo {author} {\bibfnamefont {H.}~\bibnamefont
  {Moritz}}, \bibinfo {author} {\bibfnamefont {T.}~\bibnamefont {St\"oferle}},
  \bibinfo {author} {\bibfnamefont {K.}~\bibnamefont {G\"unter}}, \bibinfo
  {author} {\bibfnamefont {M.}~\bibnamefont {K\"ohl}},\ and\ \bibinfo {author}
  {\bibfnamefont {T.}~\bibnamefont {Esslinger}},\ }\bibfield  {title} {\bibinfo
  {title} {Confinement induced molecules in a 1d fermi gas},\ }\href
  {https://doi.org/10.1103/PhysRevLett.94.210401} {\bibfield  {journal}
  {\bibinfo  {journal} {Phys. Rev. Lett.}\ }\textbf {\bibinfo {volume} {94}},\
  \bibinfo {pages} {210401} (\bibinfo {year} {2005})}\BibitemShut {NoStop}%
\bibitem [{\citenamefont {Zhou}\ and\ \citenamefont
  {Cui}(2017)}]{Zhou2017PhysRevA.96.030701}%
  \BibitemOpen
  \bibfield  {author} {\bibinfo {author} {\bibfnamefont {L.}~\bibnamefont
  {Zhou}}\ and\ \bibinfo {author} {\bibfnamefont {X.}~\bibnamefont {Cui}},\
  }\bibfield  {title} {\bibinfo {title} {Stretching $p$-wave molecules by
  transverse confinements},\ }\href
  {https://doi.org/10.1103/PhysRevA.96.030701} {\bibfield  {journal} {\bibinfo
  {journal} {Phys. Rev. A}\ }\textbf {\bibinfo {volume} {96}},\ \bibinfo
  {pages} {030701} (\bibinfo {year} {2017})}\BibitemShut {NoStop}%
\bibitem [{\citenamefont {Fonta}\ and\ \citenamefont
  {O'Hara}(2020)}]{Fonta2020PhysRevA.102.043319}%
  \BibitemOpen
  \bibfield  {author} {\bibinfo {author} {\bibfnamefont {F.}~\bibnamefont
  {Fonta}}\ and\ \bibinfo {author} {\bibfnamefont {K.~M.}\ \bibnamefont
  {O'Hara}},\ }\bibfield  {title} {\bibinfo {title} {Experimental conditions
  for obtaining halo $p$-wave dimers in quasi-one-dimension},\ }\href
  {https://doi.org/10.1103/PhysRevA.102.043319} {\bibfield  {journal} {\bibinfo
   {journal} {Phys. Rev. A}\ }\textbf {\bibinfo {volume} {102}},\ \bibinfo
  {pages} {043319} (\bibinfo {year} {2020})}\BibitemShut {NoStop}%
\bibitem [{\citenamefont {Chang}\ \emph {et~al.}(2020)\citenamefont {Chang},
  \citenamefont {Senaratne}, \citenamefont {Cavazos-Cavazos},\ and\
  \citenamefont {Hulet}}]{Chang2020PhysRevLett.125.263402}%
  \BibitemOpen
  \bibfield  {author} {\bibinfo {author} {\bibfnamefont {Y.-T.}\ \bibnamefont
  {Chang}}, \bibinfo {author} {\bibfnamefont {R.}~\bibnamefont {Senaratne}},
  \bibinfo {author} {\bibfnamefont {D.}~\bibnamefont {Cavazos-Cavazos}},\ and\
  \bibinfo {author} {\bibfnamefont {R.~G.}\ \bibnamefont {Hulet}},\ }\bibfield
  {title} {\bibinfo {title} {Collisional loss of one-dimensional fermions near
  a $p$-wave feshbach resonance},\ }\href
  {https://doi.org/10.1103/PhysRevLett.125.263402} {\bibfield  {journal}
  {\bibinfo  {journal} {Phys. Rev. Lett.}\ }\textbf {\bibinfo {volume} {125}},\
  \bibinfo {pages} {263402} (\bibinfo {year} {2020})}\BibitemShut {NoStop}%
\bibitem [{\citenamefont {Marcum}\ \emph {et~al.}(2020)\citenamefont {Marcum},
  \citenamefont {Fonta}, \citenamefont {Ismail},\ and\ \citenamefont
  {O'Hara}}]{marcum2020suppression}%
  \BibitemOpen
  \bibfield  {author} {\bibinfo {author} {\bibfnamefont {A.~S.}\ \bibnamefont
  {Marcum}}, \bibinfo {author} {\bibfnamefont {F.~R.}\ \bibnamefont {Fonta}},
  \bibinfo {author} {\bibfnamefont {A.~M.}\ \bibnamefont {Ismail}},\ and\
  \bibinfo {author} {\bibfnamefont {K.~M.}\ \bibnamefont {O'Hara}},\ }\bibfield
   {title} {\bibinfo {title} {Suppression of three-body loss near a $p$-wave
  resonance due to quasi-1d confinement},\ } {\bibfield  {journal}
  {\bibinfo  {journal} {arXiv preprint arXiv:2007.15783}\ } (\bibinfo {year}
  {2020})}\BibitemShut {NoStop}%
\bibitem [{\citenamefont {Altomare}\ and\ \citenamefont
  {Chang}(2013)}]{altomare2013one}%
  \BibitemOpen
  \bibfield  {author} {\bibinfo {author} {\bibfnamefont {F.}~\bibnamefont
  {Altomare}}\ and\ \bibinfo {author} {\bibfnamefont {A.~M.}\ \bibnamefont
  {Chang}},\ } {\emph {\bibinfo {title} {One-dimensional
  superconductivity in nanowires}}}\ (\bibinfo  {publisher} {John Wiley \&
  Sons},\ \bibinfo {year} {2013})\BibitemShut {NoStop}%
\bibitem [{\citenamefont {Alexandrou}\ \emph {et~al.}(1989)\citenamefont
  {Alexandrou}, \citenamefont {Myczkowski},\ and\ \citenamefont
  {Negele}}]{Alexandrou1989PhysRevC.39.1076}%
  \BibitemOpen
  \bibfield  {author} {\bibinfo {author} {\bibfnamefont {C.}~\bibnamefont
  {Alexandrou}}, \bibinfo {author} {\bibfnamefont {J.}~\bibnamefont
  {Myczkowski}},\ and\ \bibinfo {author} {\bibfnamefont {J.~W.}\ \bibnamefont
  {Negele}},\ }\bibfield  {title} {\bibinfo {title} {Comparison of mean-field
  and exact monte carlo solutions of a one-dimensional nuclear model},\ }\href
  {https://doi.org/10.1103/PhysRevC.39.1076} {\bibfield  {journal} {\bibinfo
  {journal} {Phys. Rev. C}\ }\textbf {\bibinfo {volume} {39}},\ \bibinfo
  {pages} {1076} (\bibinfo {year} {1989})}\BibitemShut {NoStop}%
\bibitem [{\citenamefont {Hagino}\ \emph {et~al.}(2010)\citenamefont {Hagino},
  \citenamefont {Vitturi}, \citenamefont {P{\'{e}}rez-Bernal},\ and\
  \citenamefont {Sagawa}}]{Hagino2010}%
  \BibitemOpen
  \bibfield  {author} {\bibinfo {author} {\bibfnamefont {K.}~\bibnamefont
  {Hagino}}, \bibinfo {author} {\bibfnamefont {A.}~\bibnamefont {Vitturi}},
  \bibinfo {author} {\bibfnamefont {F.}~\bibnamefont {P{\'{e}}rez-Bernal}},\
  and\ \bibinfo {author} {\bibfnamefont {H.}~\bibnamefont {Sagawa}},\
  }\bibfield  {title} {\bibinfo {title} {Two-neutron halo nuclei in one
  dimension: dineutron correlation and breakup reaction},\ }\href
  {https://doi.org/10.1088/0954-3899/38/1/015105} {\bibfield  {journal}
  {\bibinfo  {journal} {Journal of Physics G: Nuclear and Particle Physics}\
  }\textbf {\bibinfo {volume} {38}},\ \bibinfo {pages} {015105} (\bibinfo
  {year} {2010})}\BibitemShut {NoStop}%
\bibitem [{\citenamefont {Guan}\ \emph {et~al.}(2013)\citenamefont {Guan},
  \citenamefont {Batchelor},\ and\ \citenamefont
  {Lee}}]{Guan2013RevModPhys.85.1633}%
  \BibitemOpen
  \bibfield  {author} {\bibinfo {author} {\bibfnamefont {X.-W.}\ \bibnamefont
  {Guan}}, \bibinfo {author} {\bibfnamefont {M.~T.}\ \bibnamefont
  {Batchelor}},\ and\ \bibinfo {author} {\bibfnamefont {C.}~\bibnamefont
  {Lee}},\ }\bibfield  {title} {\bibinfo {title} {Fermi gases in one dimension:
  From bethe ansatz to experiments},\ }\href
  {https://doi.org/10.1103/RevModPhys.85.1633} {\bibfield  {journal} {\bibinfo
  {journal} {Rev. Mod. Phys.}\ }\textbf {\bibinfo {volume} {85}},\ \bibinfo
  {pages} {1633} (\bibinfo {year} {2013})}\BibitemShut {NoStop}%
\bibitem [{\citenamefont {Cui}(2016)}]{Cui2016PhysRevA.94.043636}%
  \BibitemOpen
  \bibfield  {author} {\bibinfo {author} {\bibfnamefont {X.}~\bibnamefont
  {Cui}},\ }\bibfield  {title} {\bibinfo {title} {Universal one-dimensional
  atomic gases near odd-wave resonance},\ }\href
  {https://doi.org/10.1103/PhysRevA.94.043636} {\bibfield  {journal} {\bibinfo
  {journal} {Phys. Rev. A}\ }\textbf {\bibinfo {volume} {94}},\ \bibinfo
  {pages} {043636} (\bibinfo {year} {2016})}\BibitemShut {NoStop}%
\bibitem [{\citenamefont {Tajima}\ \emph
  {et~al.}(2021{\natexlab{b}})\citenamefont {Tajima}, \citenamefont {Tsutsui},
  \citenamefont {Doi},\ and\ \citenamefont
  {Iida}}]{Tajima2021PhysRevA.104.023319}%
  \BibitemOpen
  \bibfield  {author} {\bibinfo {author} {\bibfnamefont {H.}~\bibnamefont
  {Tajima}}, \bibinfo {author} {\bibfnamefont {S.}~\bibnamefont {Tsutsui}},
  \bibinfo {author} {\bibfnamefont {T.~M.}\ \bibnamefont {Doi}},\ and\ \bibinfo
  {author} {\bibfnamefont {K.}~\bibnamefont {Iida}},\ }\bibfield  {title}
  {\bibinfo {title} {Unitary $p$-wave fermi gas in one dimension},\ }\href
  {https://doi.org/10.1103/PhysRevA.104.023319} {\bibfield  {journal} {\bibinfo
   {journal} {Phys. Rev. A}\ }\textbf {\bibinfo {volume} {104}},\ \bibinfo
  {pages} {023319} (\bibinfo {year} {2021}{\natexlab{b}})}\BibitemShut
  {NoStop}%
\bibitem [{\citenamefont {Guo}\ and\ \citenamefont
  {Tajima}(2022)}]{guo2022stability}%
  \BibitemOpen
  \bibfield  {author} {\bibinfo {author} {\bibfnamefont {Y.}~\bibnamefont
  {Guo}}\ and\ \bibinfo {author} {\bibfnamefont {H.}~\bibnamefont {Tajima}},\
  }\bibfield  {title} {\bibinfo {title} {Stability against three-body
  clustering in one-dimensional spinless $p$-wave fermions},\ }\href
  {https://doi.org/10.1103/PhysRevA.106.043310} {\bibfield  {journal} {\bibinfo
   {journal} {Phys. Rev. A}\ }\textbf {\bibinfo {volume} {106}},\ \bibinfo
  {pages} {043310} (\bibinfo {year} {2022})}\BibitemShut {NoStop}%
\bibitem [{\citenamefont {Sekino}\ and\ \citenamefont
  {Nishida}(2021)}]{Sekino2021Phys.Rev.A103.043307}%
  \BibitemOpen
  \bibfield  {author} {\bibinfo {author} {\bibfnamefont {Y.}~\bibnamefont
  {Sekino}}\ and\ \bibinfo {author} {\bibfnamefont {Y.}~\bibnamefont
  {Nishida}},\ }\bibfield  {title} {\bibinfo {title} {Field-theoretical aspects
  of one-dimensional bose and fermi gases with contact interactions},\ }\href
  {https://doi.org/10.1103/PhysRevA.103.043307} {\bibfield  {journal} {\bibinfo
   {journal} {Phys. Rev. A}\ }\textbf {\bibinfo {volume} {103}},\ \bibinfo
  {pages} {043307} (\bibinfo {year} {2021})}\BibitemShut {NoStop}%
\bibitem [{\citenamefont {Thomas}(1935)}]{Thomas1935Phys.Rev.47.903--909}%
  \BibitemOpen
  \bibfield  {author} {\bibinfo {author} {\bibfnamefont {L.~H.}\ \bibnamefont
  {Thomas}},\ }\bibfield  {title} {\bibinfo {title} {The interaction between a
  neutron and a proton and the structure of ${\mathrm{h}}^{3}$},\ }\href
  {https://doi.org/10.1103/PhysRev.47.903} {\bibfield  {journal} {\bibinfo
  {journal} {Phys. Rev.}\ }\textbf {\bibinfo {volume} {47}},\ \bibinfo {pages}
  {903} (\bibinfo {year} {1935})}\BibitemShut {NoStop}%
\bibitem [{\citenamefont {He}\ \emph {et~al.}(2019)\citenamefont {He},
  \citenamefont {Tian}, \citenamefont {Pekker},\ and\ \citenamefont
  {Mong}}]{He2019Phys.Rev.B100.201101}%
  \BibitemOpen
  \bibfield  {author} {\bibinfo {author} {\bibfnamefont {Y.}~\bibnamefont
  {He}}, \bibinfo {author} {\bibfnamefont {B.}~\bibnamefont {Tian}}, \bibinfo
  {author} {\bibfnamefont {D.}~\bibnamefont {Pekker}},\ and\ \bibinfo {author}
  {\bibfnamefont {R.~S.~K.}\ \bibnamefont {Mong}},\ }\bibfield  {title}
  {\bibinfo {title} {Emergent mode and bound states in single-component
  one-dimensional lattice fermionic systems},\ }\href
  {https://doi.org/10.1103/PhysRevB.100.201101} {\bibfield  {journal} {\bibinfo
   {journal} {Phys. Rev. B}\ }\textbf {\bibinfo {volume} {100}},\ \bibinfo
  {pages} {201101} (\bibinfo {year} {2019})}\BibitemShut {NoStop}%
\bibitem [{\citenamefont
  {Schollw\"ock}(2005)}]{Schollwoeck2005Rev.Mod.Phys.77.259--315}%
  \BibitemOpen
  \bibfield  {author} {\bibinfo {author} {\bibfnamefont {U.}~\bibnamefont
  {Schollw\"ock}},\ }\bibfield  {title} {\bibinfo {title} {The density-matrix
  renormalization group},\ }\href {https://doi.org/10.1103/RevModPhys.77.259}
  {\bibfield  {journal} {\bibinfo  {journal} {Rev. Mod. Phys.}\ }\textbf
  {\bibinfo {volume} {77}},\ \bibinfo {pages} {259} (\bibinfo {year}
  {2005})}\BibitemShut {NoStop}%
\bibitem [{\citenamefont {S{\'e}n{\'e}chal}(2004)}]{senechal2004introduction}%
  \BibitemOpen
  \bibfield  {author} {\bibinfo {author} {\bibfnamefont {D.}~\bibnamefont
  {S{\'e}n{\'e}chal}},\ }\bibfield  {title} {\bibinfo {title} {An introduction
  to bosonization},\ }in\  {\emph {\bibinfo {booktitle}
  {Theoretical Methods for Strongly Correlated Electrons}}}\ (\bibinfo
  {publisher} {Springer},\ \bibinfo {year} {2004})\ pp.\ \bibinfo {pages}
  {139--186}\BibitemShut {NoStop}%
\end{thebibliography}

%

\end{CJK}
\end{document}